\documentclass[preliminary,copyright,creativecommons]{eptcs}
\providecommand{\event}{MARS 2022} 
\usepackage{xspace}
\usepackage{xcolor}
\usepackage{afterpage}
\usepackage{graphicx}
\usepackage{amsfonts,amsmath,amssymb}
\usepackage{enumitem}
\setitemize{noitemsep,topsep=0pt,parsep=0pt,partopsep=0pt,leftmargin=0.7cm}
\setenumerate{noitemsep,topsep=0pt,parsep=0pt,partopsep=0pt,leftmargin=0.7cm}
\usepackage{wrapfig}

\definecolor{backgroundColour}{rgb}{0.95,0.95,0.92}

\usepackage{hyperref}

\newcommand{\F}{\texttt{<>}}
\newcommand{\G}{\texttt{[]}}

\newcommand{\ie}{i.e.\ }
\newcommand{\eg}{e.g.\ }

\newcommand{\uppaal}{{\sc Uppaal}\xspace}

\newcommand{\ignore}[1]{}

\newcommand{\myparagraph}[1]{\vspace{-4.5mm}\paragraph{#1}}
\newcommand{\hellomsg}{{\scshape Hello} message\xspace}
\newcommand{\hellomsgs}{{\hellomsg}s\xspace}
\newcommand{\dbdmsg}{{\scshape Dbd} message\xspace}
\newcommand{\dbdmsgs}{{\dbdmsg}s\xspace}
\newcommand{\lsadv}{{\scshape Lsa}\xspace}
\newcommand{\lsadvs}{{\lsadv}s\xspace}
\newcommand{\lsrmsg}{{\scshape Lsr} message\xspace}
\newcommand{\lsrmsgs}{{\lsrmsg}s\xspace}

\newcommand{\lsumsg}{{\scshape Lsu} message\xspace}
\newcommand{\lsumsgs}{{\lsumsg}s\xspace}

\newcommand{\Down}{\textsf{Down}\xspace}
\newcommand{\Init}{\textsf{Init}\xspace}
\newcommand{\ExStart}{\textsf{ExStart}\xspace}
\newcommand{\Exchange}{\textsf{Exchange}\xspace}
\newcommand{\Loading}{\textsf{Loading}\xspace}
\newcommand{\Full}{\textsf{Full}\xspace}

\title{Advanced Models for the OSPF Routing Protocol}
\def\titlerunning{Advanced Models for OSPF}
\author{Courtney Darville
\institute{Data61, CSIRO, Sydney, Australia}
\institute{University of New South Wales, Sydney, Australia}
\email{C.Darville@unsw.edu.au}
\and
Peter H\"ofner
\institute{School of Computing\\ANU, Canberra, Australia}
\email{Peter.Hoefner@anu.edu.au}
\and
Franc Ivankovic
\institute{University of Trento, Trento, Italy}
\email{francekivankovic@gmail.com}
\and
Adam Pam
\institute{Macquarie University, Sydney, Australia}
\email{Adam.Pam@students.mq.edu.au}
}

\def\authorrunning{C. Darville, P. H\"ofner, I.  Ivankovic \& A. Pam}
\begin{document}
\maketitle
\vspace{-2mm}
\begin{abstract}
	We present two formal models for the OSPF routing protocol, 
	designed for the model checker \uppaal.
	The first one is an optimised model of an existing model that
	allows to check larger network topologies.
	The second one is a specialised model for adjacency building, a complex subprocedure of OSPF,
     which is not part of any existing model and which is known to be vulnerable 
	to cyber attacks. We illustrate how both models can be used to discover vulnerabilities in routing protocols.
\end{abstract}
\vspace{-2mm}

\section{Introduction\label{sec:intro}}

The Open Shortest Path First (OSPF) protocol~\cite{rfc2328} is a widely used, proactive, link-state routing protocol that distributes routing information throughout a single autonomous system.  Similarly to many other routing protocols, it aims to establish shortest paths between network routers while keeping network overhead --- messages carrying routing information rather than user data --- to a minimum. 
Although OSPF, as many other routing protocols, is based on `simple' algorithms such as Dijkstra's shortest path algorithm~\cite{Dijkstra59},
it seems incredibly hard to ensure that the protocol is functionally correct, see \eg \cite{MK10,MSWIM13}.

To allow formal analysis of OSPF, three detailed formal models have been presented in~\cite{DHW20}:\footnote{Some other models of OSPF exist~\cite{NakiblyEtAl14,Malik}; see \autoref{sec:related} and \cite{DHW20} for a discussion.}
two are formalised in the timed process algebra T-AWN~\cite{ESOP16}, which is not only tailored to routing protocols, but also specifies protocols in pseudo-code that is easily readable. The difference between the two models lies in the level of detail. 
The more abstract model is the basis for the third, which presents OSPF as a network of timed automata that can be executed in the model checker  \uppaal~\cite{LPY97,uppaal04}.

The process-algebraic models do not allow automatic analysis of the protocol, for there is currently no tool support available for T-AWN. 
\uppaal,  however, provides several tools, from simulation to model checking. Unfortunately, the current model faces a combinatorial blow up of the state space that must be addressed before automatic analysis of the protocol becomes feasible. In fact, the current model is already infeasible when analysing networks consisting of three nodes only. 

Based on the existing \uppaal-model, we advance the modelling efforts for OSPF:
\begin{enumerate}
\item We develop a \uppaal model that is behaviourally equivalent to the model presented in~\cite{DHW20} but with a much smaller state space.
\item We develop a new model for adjacency building, a subprocedure of OSPF, which so far has been ignored in formal models.
\item We develop flexible models of adversaries that cover many different attack scenarios. Although the modelling is straightforward, it is extremely powerful and can be used to discover vulnerabilities. 
\item We illustrate how this adversarial model can be used in combination with both new models to discover vulnerabilities in OSPF.
\end{enumerate}
\pagebreak

\section{The Open Shortest Path First (OSPF) Protocol\label{sec:ospf}}

The \emph{Open Shortest Path First (OSPF) protocol}~\cite{rfc2328} is a link-state protocol that falls into the group of interior gateway protocols, operating within a single autonomous system. As with any link-state protocol, OSPF routers (nodes in a network) exchange topological information about one-hop links with their neighbours, \ie the nodes within transmission range. 
That information is distributed through the network so that (eventually) every router has a complete picture of all available links. 
This knowledge is used to calculate the shortest/best route between any two nodes, using a variant of Dijkstra's algorithm~\cite{Dijkstra59}. 

To discover immediate (one-hop) neighbours, OSPF uses \emph{H\scalebox{0.75}{ELLO} messages}, which are broadcast periodically by all nodes. 
\hellomsgs consist of the sender's unique identification (ID) and a list of all known one-hop neighbours of the sender, \ie  nodes from which the sender has received \hellomsgs. 
Upon receipt of a \hellomsg, a node updates its own data structures accordingly. 
The receiver consequently drops that \hellomsg, \ie \hellomsgs are not forwarded.

A node distributes information about its known connections using \emph{Link State Advertisements} ({\lsadvs). 

A node stores data concerning connections between other nodes in their \emph{Link-State Database (LSDB)}. 
This database represents the router's current view of the network topology.
It contains the most recently received \lsadvs from each unique originator. 
Next to the LSDB, every node maintains a list of discovered neighbours, with whom routing information may be exchanged.

\hellomsgs (or the lack thereof) are also used to determine whether nodes have become inactive or lost connectivity.
When a node receives the first \hellomsg from a neighbour, it learns of its existence. 
If the node finds its own ID listed within that \hellomsg\ --- meaning that the neighbour is aware of the node --- it checks whether it needs to form an adjacency with that~neighbour. 

The term \emph{adjacency} describes the relationship between two neighbouring nodes that exchange all of their topological information --- the content of their LSDBs. 
After information has been shared between two adjacent nodes, they have an identical understanding of the entire network. 
Nodes not forming an adjacency do not share this information and hence network overhead is reduced.
When two nodes recognise that they need to form an adjacency, the node with the larger node ID becomes the \emph{master}, and the other becomes the \emph{slave}.  
It is the master that initiates the exchange of data by sending the necessary information using \emph{Database Description (D\scalebox{0.75}{BD}) messages}. 
After that, the two nodes exchange a sequence of \dbdmsgs. 
We describe further details about adjacencies in~\autoref{sec:adjacency}.

Once a node has received a full description of the other node's LSDB, it compares that description with its own LSDB. 
To resolve inconsistencies between the two, it sends \emph{Link State Request ({L\scalebox{0.75}{SR}}) messages}, asking for \lsadvs containing the newest available information.
\lsrmsgs are answered by \emph{Link State Update (L\scalebox{0.75}{SU}) messages}; they contain the requested \lsadvs.
Each receipt of an \lsumsg is acknowledged by a \emph{Link State Acknowledgement (L\scalebox{0.75}{SACK}) message}.

\emph{Designated routers} are used in combination with adjacencies to further reduce network traffic.  
As they are not crucial to the understanding of the main functionality of OSPF, we omit the details.

\section{Modelling OSPF in \uppaal\label{sec:ospfuppaal}}
One of our aims is to analyse OSPF automatically, both with regards to functional correctness and to discovery of vulnerabilities. A  standard tools for analysing and verifying systems automatically is model checking.
As there are existing models for \uppaal, we have chosen to stick with that model checker.

\uppaal~\cite{uppaal04,LPY97} is an established model checker, which is frequently used for protocol verification, \eg \cite{BOG02,WPP04,RA04,FHM07,TACAS12}.
\uppaal analyses \emph{networks of timed automata},
with clocks supporting the modelling and the analysis of temporal aspects. 
It provides two synchronisation mechanisms: \emph{binary} and \emph{broadcast channels}. 
In the setting of routing protocols these usually translate to unicast and broadcast communication; 
the latter presenting the transmission of messages to \emph{all} neighbours of a node.
Hence, when using \uppaal's broadcast mechanism one must consider the network topology to determine
the exact set of nodes that are able to receive messages (see below).
\uppaal also provides common data structures, such as arrays, and a C-like programming language to define updates on these data structures. 

\myparagraph{Networks of Timed Automata}
The state of the system  is determined, in part, by the values of data variables that can be either local or shared between automata. We assume a data structure with several types, variables ranging over these types, operators and  predicates. Common Boolean and arithmetic expressions are used to denote data values and statements about them.

The automata are extended with clock variables. \uppaal uses a dense-time model where a clock variable evaluates to a real number. All the clocks progress synchronously. 

Each automaton is a graph, with locations, and edges between
locations. Every edge has a guard, optionally a synchronisation label,
and an update, which allows  local and global data structures to be updated. Synchronisation occurs via so-called channels; for each
channel $a$ there is one label $a!$ to denote the sender, and $a?$ to
denote the receiver. Transitions without labels are internal; all other transitions use one of the two types of synchronisation.

\myparagraph{Synchronisation}
In \emph{binary handshake} synchronisation, an automaton having an edge with a label that has the suffix~$!$ synchronises with another automaton with an edge having the same label with a $?$-suffix. These two transitions may synchronise if and only if both guards are true in the current state of the system. When the transition is taken, both locations change, and the updates will be applied to the state variables; first the updates on the $!$-edge, then the updates on the $?$-edge. If there is more than one possible pair, then the transition is selected non-deterministically.

In \emph{broadcast} synchronisation, one automaton with a $!$-labelled edge synchronises
with the set of other automata that all have a matching edge with a $?$-label. The
initiating automaton can change its location, and apply its update, if the guard on its edge
evaluates to true. It does not require a second synchronising automaton. Automata with a
matching $?$-labelled edge have to synchronise if their guard is currently true. They change their location and update the system state. The automaton with the $!$-edge will update the state first, followed by the other automata in some lexicographic order. If more than one automaton can initiate a transition on an $!$-edge, the choice will be made non-deterministically.

\myparagraph{Network Communication}
To model the network mechanisms of unicast, broadcast and multicast --- sending to a dedicated set of neighbours --- we combine \uppaal's synchronisation with guards. 
Message contents are exchanged via global data using \uppaal's update mechanism.
\begin{figure}[t]
\centering
\vspace{-2mm}
\begin{tabular}[t]{r@{\hspace{17mm}}l}
\includegraphics[scale=.95]{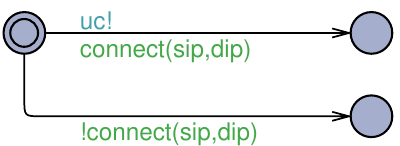} &
\raisebox{23pt}{\includegraphics[scale=.95]{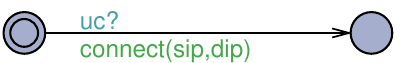}}\\[1mm]
\includegraphics[scale=.95]{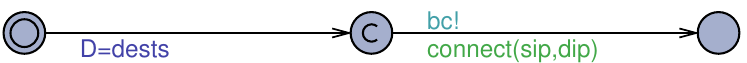}&
\raisebox{-6pt}{\includegraphics[scale=.95]{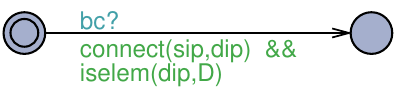}}
\end{tabular}
\caption[]{Model Network Message Passing: unicast and multicast\label{fig:uppaal}}
\vspace{-2mm}
\end{figure}

For communication within routing protocols, 
unicast is a node-to-node communication with an in-built  acknowledgment-of-receipt mechanism, which is implemented at a lower layer in the network stack.\footnote{Usually implemented by the link layer of relevant standards such as IEEE 802.11~\cite{IEEE80211}.}
More formally, unicast in networking means that a node \textsf{sip} can only send to another node \textsf{dip}
if (a) both nodes are within transmission range of each other, (b) \textsf{dip} is ready to receive and (c) after attempting to 
send a message, the node \textsf{sip} is aware of whether transmission was successful. 
The corresponding model in \uppaal is depicted on the top of \autoref{fig:uppaal}.
It uses a unicast channel \textsf{uc}.
The network topology is modelled by a Boolean predicate \textsf{connect}.
The sender has two successors, one for successful transmission --- as we do not model 
lossy channels, this can be modelled as \textsf{connect(sip,dip)} --- and one for failed transmission (\textsf{!connect}), 
which models the actions of a failed sending attempt.
The receiving edge is straight forward. 

Broadcast in networking means sending a message to all other network nodes within transmission range.
Similarly to unicast, this is modelled by \uppaal's broadcast mechanism --- the broadcast channel is named \textsf{bc} --- in combination with the 
predicate \textsf{connect}. However, the model has to ensure that all nodes within transmission range are 
in a state where messages can be received.

A restricted version of broadcast is multicast: a node \textsf{sip} tries
to transmit a message to destinations \textsf{dests}, and proceeds
regardless of whether any of the transmissions is successful.
The corresponding modelling in \uppaal is shown on the bottom of  \autoref{fig:uppaal}.
Here, \textsf{D} is a global data structure, maintaining the set of intended destinations. 
Before the actual message transfer occurs, the sender \textsf{sip} has to update \textsf{D}. 
By doing so, nodes within transmission range of \text{sip} can determine whether they are intended recipients, 
using the Boolean function \textsf{iselem}.

\section{Optimising a Model of OSPF\label{sec:modeli}}
Drury et al.~\cite{DHW20} have created a \uppaal-model that closely follows the OSPF specification \cite{rfc2328};~they also take into account the amendments described in~\cite{rfc5340}.  In that model, each network router is modelled by two automata: one describing the core behaviour of OSPF and the other a queue for outgoing messages.

\begin{figure}[t]
\centering
\vspace{0mm}
\includegraphics[width=1.0\textwidth]{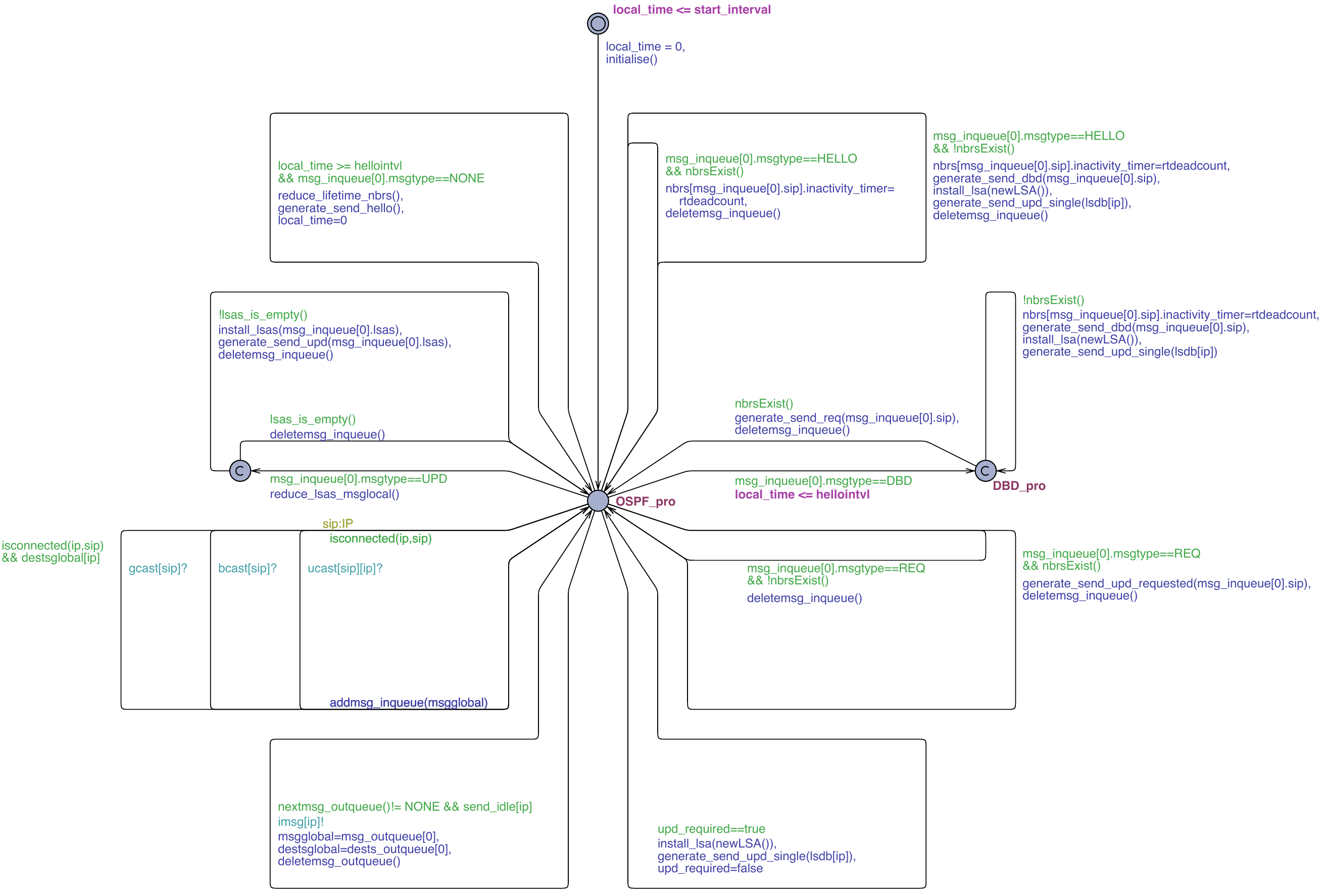}
\vspace{-7mm}
\caption[]{An OSPF router as a timed automaton\footnotemark \label{fig:ospf}}
\vspace{0mm}
\end{figure}
\footnotetext{\label{note1}The diagrams are intended to show the models' complexity, not to be ‘read’.\newline\mbox{}\hspace{5mm}
All models are available at \url{http://mars-workshop.org}.
}
The main automaton (\autoref{fig:ospf}) is built around a `central' location \textsf{OSPF\_pro}.
This location models an idle state of the protocol --- the protocol can stay there indefinitely. 
As soon as a message is received, the automaton identifies the type of message and acts accordingly.
For example, the two transitions in the upper right of \autoref{fig:ospf} model the receipt of an incoming \hellomsg:
the automaton checks whether the content of the message shows the node's ID (see \autoref{sec:ospf}) and takes the corresponding transition. 
The automaton uses a clock to send out \hellomsgs periodically (upper left of \autoref{fig:ospf}).
The receipt of messages does not take time; only the transmission of messages does. 
When the protocol needs to send a message, this automaton creates the message and passes it, via unicast-synchronisation, to another automaton that models a queue for outbound messages. The queue manages the delivery of the message to corresponding recipients. 
Each node maintains a data structure for LSDBs, two independent arrays of messages to be sent (as part of the queue automaton) and to be received (part of the main automaton), 
as well as two clocks, one for each automaton. 
Each router makes use of four different channels, one for internal communication, one for broadcast, one for unicast, and one for multicast.

As often occurs, this model faces a combinatorial blow up of the state space
that must be addressed to analyse the protocol automatically.
The state-space explosion is caused by multiple factors:
 \begin{enumerate}[nosep]
 	\item Each additional router adds two automata to the network of timed automata.
 	\item Each additional router adds two clocks to the model, which exponentially increases the complexity.
 	\item The underlying data structures, messages and message queues significantly add to the state space.
 \end{enumerate}
 As a result, checking even simple properties, such as deadlock, fails already on topologies of size three.\footnote{Deadlock checking failed on a machine with an AMD Ryzen 7 CPU and 64GB RAM.}

To overcome (some of) the problems regarding the state space, we create a single automaton describing the entire network. It is depicted in \autoref{fig:composer}.
 \begin{figure}[t]
 \centering
 \includegraphics[width=1\textwidth]{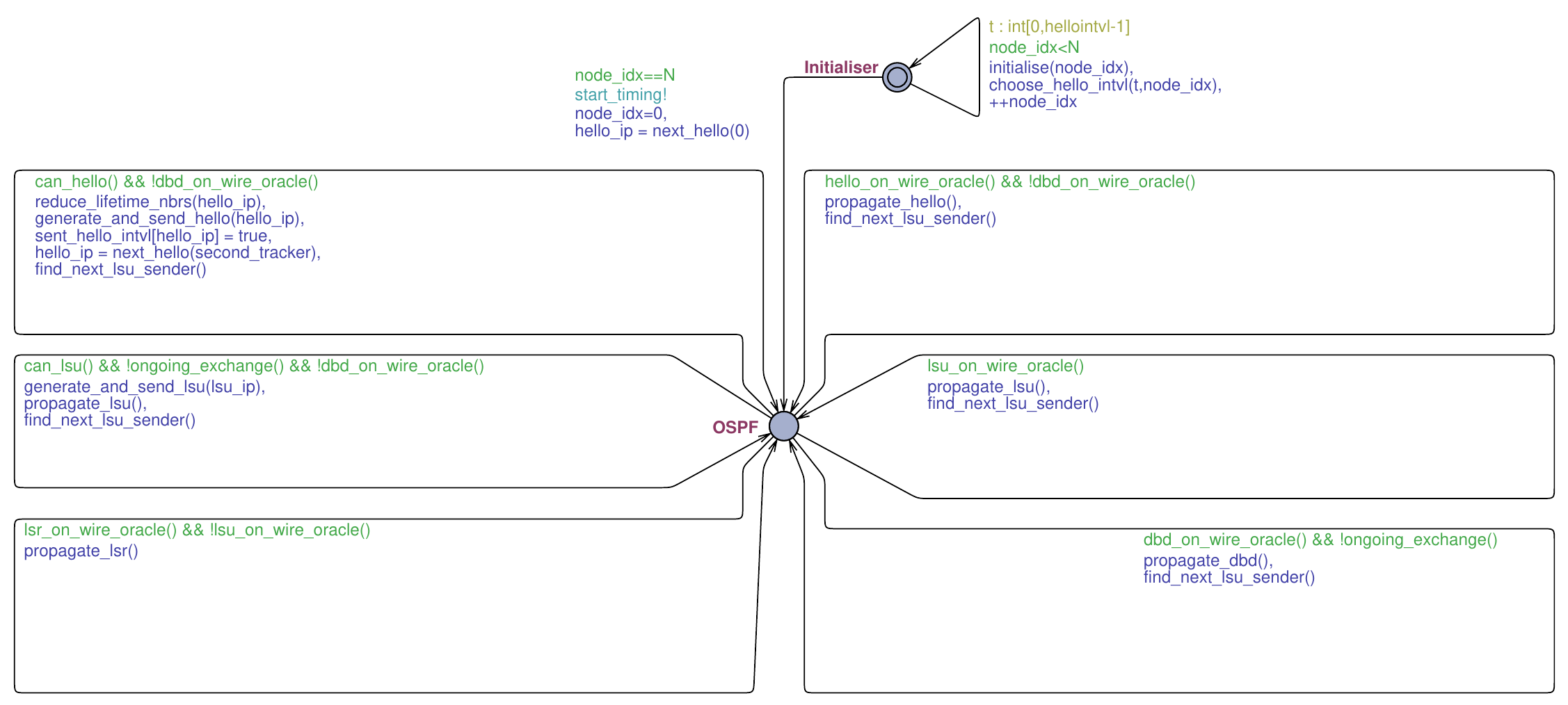}
 \caption[]{A single automaton for OSPF\label{fig:composer}}
 \end{figure}
  
This single automaton models the nondeterministic interaction between routers, rather than actual routers. 
Data structures of the individual nodes are organised into multidimensional arrays.
This allows a single automaton to gain all knowledge of the network and it is no longer necessary to model the concrete content of OSPF messages. We instead model message transmission as a function that extracts information from the sender's `local' data structures, such as a the LSDB, and updates the `local' data structure of the receiver(s) directly. 

As the entire model is assembled into a single automaton.
We summarise some advantages of our newly developed model, compared to the original model:
 \begin{enumerate}[nosep]
 	\item The content of messages are not modelled; hence a significant reduction in the complexity of the  required data structures.
 	\item Only one clock is required rather than $2\mathop\cdot N$, where $N$ is the number of routers in a network; the model requires an additional small data structure to manage the periodic sending of \hellomsgs.
 	\item Message queues and message propagation delays are removed; the model provides similar behaviour through a choice of nondeterministic transitions.
 \end{enumerate}

A simple automaton is added alongside the main automaton to monitor the model's clock and notify the main automaton when \hellomsgs need to be sent for each router. Although we did not prove the relationship to the original model of OSPF, the model is carefully designed that it is behaviourally equivalent to the original one. 

In a further abstraction, we can replace the (dense-time) clock by a bounded integer that takes over the periodic sending of \hellomsgs.
We assume that this model is behaviourally equivalent with regards to message sending and data structures, but obviously not regarding timing properties.

\section{Adjacency Building\label{sec:adjacency}}

The original model of OSPF, as well as the model described above, capture all but one core functionality of OSPF. The missing functionality, adjacency building, is an independent subroutine with no immediate effect on other functionality and 
hence is usually abstracted away.
As discussed in \autoref{sec:ospf}, \emph{adjacency building} describes the activities involved when two neighbouring nodes exchange all of their topological information -- the content of their LSDBs. 
After the exchange, the nodes have an identical understanding of the entire network. 

In this section, we present a new, detailed model of adjacency building.
When nodes have been assigned to form an adjacency, they progress through the following six \emph{neighbour states}:
\begin{figure}[th]
\centering
\includegraphics[width=1\textwidth]{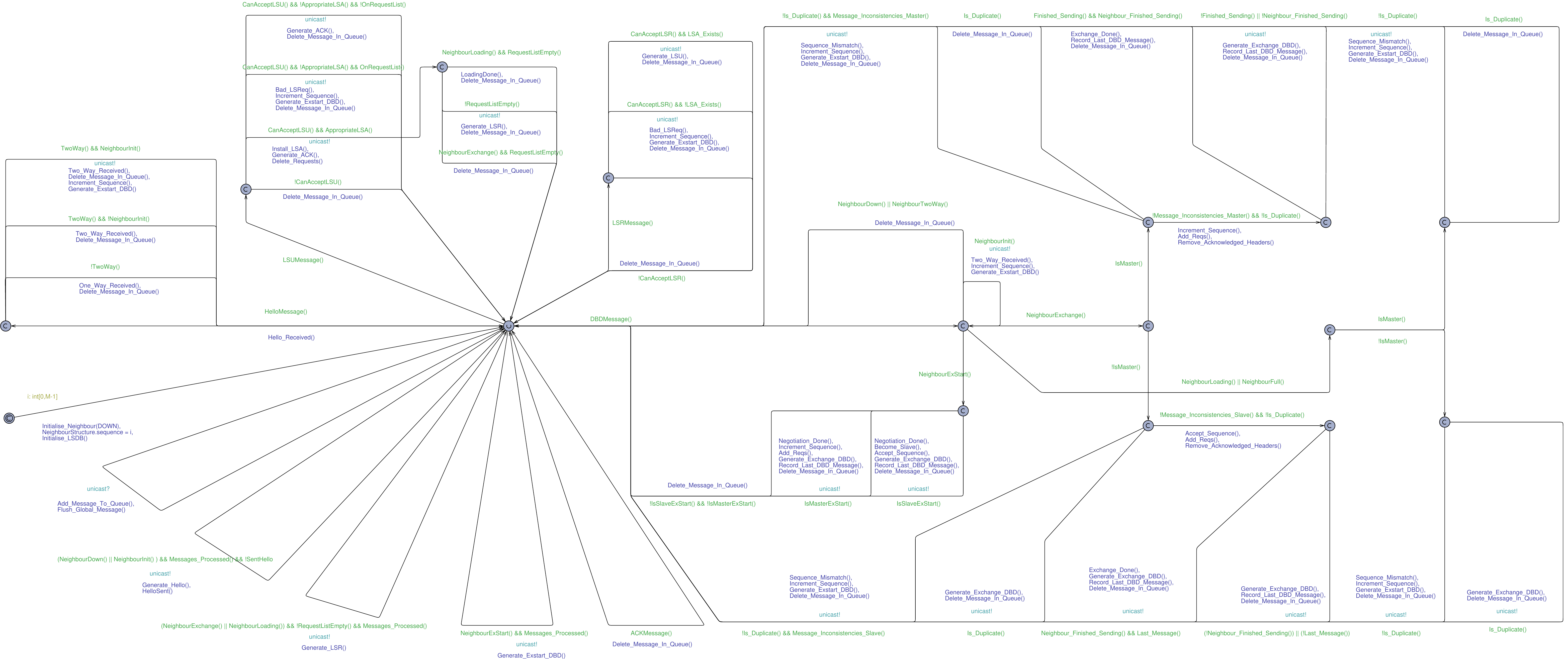}
\caption[]{Adjacency Building$^{\ref{note1}}$\label{fig:adjacency}}
\end{figure}
\begin{description}[nosep]
	\item[\Down:] No information has been exchanged between the two nodes.
	\item[\Init:]
	     A participating node has recently received a \hellomsg from a neighbour. However, bidirectional communication is 
		not confirmed yet -- the node's own IP is not part of the \hellomsg.
	\item[\ExStart:]
		This is the first step in the adjacency-establishment procedure; it establishes the master/slave relationship between the nodes.
	\item[{\smash{\Exchange:}}] The master/slave relationship has been agreed upon, and nodes send \dbdmsgs summarising their LSDBs.
	\item [{\smash{\Loading:}}] Nodes compare the full description of their neighbour's LSDB with their local data. 
		Missing information is requested via \lsrmsgs.
	\item[\Full:] All requested \lsadvs have been received.
\end{description}
\noindent A node’s neighbour state determines which messages it will send to its neighbour, as well as how it responds to messages received.


\begin{wrapfigure}[18]{r}{.47\textwidth}
	\hspace{0mm}\includegraphics[width=.48\textwidth]{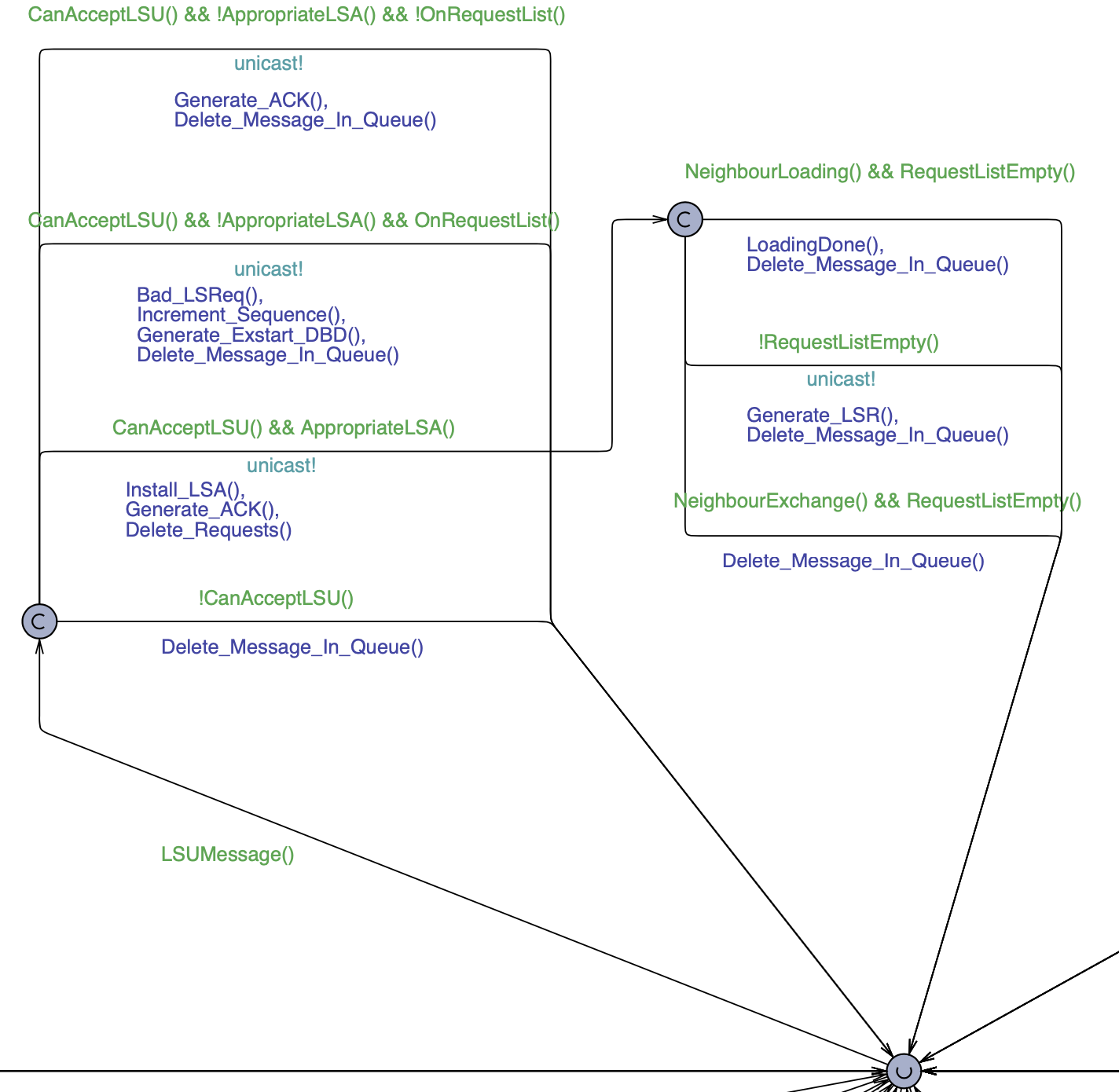}
	\vspace{-4.5mm}
	\caption{Receiving of \lsumsgs\label{fig:LSU}}
\end{wrapfigure}
Our adjacency model, depicted in \autoref{fig:adjacency}, models the entire adjacency process from the \Down state to the \Full state. 
Similar to the original model~\cite{DHW20}, the adjacency model uses one automaton per node.
In contrast to the previous model, we do not encounter problems with the state space as adjacency building is an activity between two nodes (two automata). 
The model closely follows the OSPF standard. Due to its complexity -- it would take a lot of space to explain all details of the adjacency-building process, so they are omitted.  The model can be found at \url{http://mars-workshop.org} and should be self-explanatory. 
The model does not include a clock, and instead is forced to make progress by using urgent and committed states within the automata.%
\footnote{For a formal definition of urgent and committed locations see~\cite{LPY97}.} The automata exchange messages using binary synchronisation.
To provide some insight in our model, we provide details on the handling of \lsumsgs received (\autoref{fig:LSU}).

The transition from the central state to a committed state ensures that the message received is indeed an \lsumsg. 
There are four possibilities as to how this message can be handled. 
The transition guarded by \textsf{!CanAcceptLSU()} corresponds to the case where the node is not in the state \Exchange, \Loading or \Full, and hence any incoming \lsumsg is ignored and deleted.
The top two transitions correspond to cases where a node can accept an \lsumsg, but the received message contains outdated or incorrect information.
The final transition models the default case where the content of the \lsumsg is  more recent than the information stored in the node's LSDB. 
In this case, the node updates its LSDB and \lsadv  request list, and generates an acknowledgement message in response.

To validate the correctness of our model, we run some sanity checks. 
Among others we check that adjacencies are always completely established, \ie both nodes reach the state \Full, and that  LSDBs are identical when the adjacency is established. 
Using \uppaal's CTL syntax, these properties correspond~to 
\[{\begin{array}{r@{}l}
\textsf{A\F(}&\textsf{n1.NeighbourState == FULL \&\& n2.NeighbourState == FULL)}\ \  \text{and}\\
\textsf{A\G(}&\textsf{(n1.NeighbourState == FULL \&\& n2.NeighbourState == FULL)}\textsf{ imply LSDB\_Sync())}
\end{array}
}\]%
\noindent Here \textsf{n1} and \textsf{n2} are the two nodes forming an adjacency, and  \textsf{LSDB\_Sync()} is a function that checks LSDBs for synchronicity.
The CTL formula \textsf{A\F$\varphi$} is satisfied if $\varphi$ holds on some state along all paths, 
and  \textsf{A\G$\varphi$} is satisfied if $\varphi$ holds on all states along all paths (\eg \cite{Emerson91}).
We are able to verify these properties for a number of different starting configurations of the node’s LSDBs. 

\section{Modelling Adversaries\label{sec:adversaries}}
Over the  last decades, model checking has become one of the standard tools for analysing and verifying systems,
with the main focus on functional correctness, \ie checking whether a system acts as expected.

Although correctness proofs are important, they are only half of the story.
With systems connected more than ever, detecting and analysing vulnerabilities of systems are of equal importance. 
In the general area of security protocols, including cryptographic protocols, modelling adversaries and finding vulnerabilities using model checking techniques are common~\cite{MSCB13,Blanchet16}. With very few exceptions, \eg \cite{HLS99}, finding vulnerabilities in systems, such as routing protocols, using model checkers is uncommon.

In this section, we present first attempts to discover vulnerables, using model checking.
When modelling routing attacks we distinguish two different concepts: \emph{attack capabilities} and \emph{attack goals}. Attack capabilities define the behaviour of the adversary. 
They include properties such as the number of malicious nodes and connectivities (\eg \cite{NKGB12}).
In contrast, an attack goal describes the specific aim of an attack.
This includes general properties such as delivery failure and routing loops, as well as protocol-specific targets such as the destruction of adjacencies.
The consequences of (successful) attacks are sometimes also analysed. 
For example, a routing-loop attack yields packet loss. 

\subsection{Attack Capabilities\label{ssec:cap}}
We  model capabilities as one or more separate timed automata.
This modular approach allows us to combine the same attacker model with different formalisations of OSPF, or even other routing protocols.

\myparagraph{Number of Malicious Nodes}
The \emph{number of malicious nodes} in the network is one of the capabilities. 
In our setting, each malicious node (attacker) is modelled by a separate automaton executed in parallel to the model 
of the routing protocol. For the moment, we restrict ourselves to one malicious node only.

\myparagraph{Connectivity} There exist two reasonable scenarios: 
(a)~a malicious node can only send to its immediate neighbours, similar to a node running an uncorrupted version of OSPF, and 
(b)~a node can inject messages to all nodes in the network, which can be achieved in real networks (\eg~\cite{NKGB12}).
In both models of OSPF, we characterise connectivity as a Boolean matrix, or equivalently with a Boolean predicate \textsf{connect}, see \autoref{sec:ospfuppaal}. 
When modelling capability (a), no change is required.  For option (b), a minor modification of the predicate \textsf{connect} suffices.
Technically, the relation is not symmetric any longer.

\myparagraph{Time of the Attack} 
The third capability is the \emph{time point an attack begins}.  It determines at what point during the lifetime of the network a malicious actor performs the take over of a router and begins an attack. 
Although this point in time is arbitrary, we only consider two scenarios: 
(a) the node is already malicious when the protocol is initiated, and 
(b) the attacker gains access to a node after OSPF reaches a steady state, \ie all adjacencies are established and all nodes have the same view of the network topology. 

Other possible time points,  \eg the attack starts during initialisation of the protocol, are not considered since the likelihood of exploitation during these time periods is basically zero.

\myparagraph{Adversaries' Activities}
The final, and arguably most important, capability describes the \emph{activities} an adversary can perform. 
They are manifold, sometimes protocol-specific, and include the following:
\begin{itemize}[nosep]
\item type of OSPF messages ({\scshape Dbd}, {\scshape Lsr}, {\scshape Lsu}, \dots)  the attacker is allowed to send;
\item components of the message which can be manipulated; \eg the attacker can  `only'  change the content of a message or 
  it can pretend to send the message on behalf of some other node (false identity);
\item the order of messages sent; \dots
\end{itemize}
For our initial experiments, described below, we use an `all mighty' adversary that can inject any type of message with arbitrary content at any time. We model this behaviour as a nondeterministic automaton.

 \begin{figure}[t]
 \centering
 \includegraphics[width=1\textwidth]{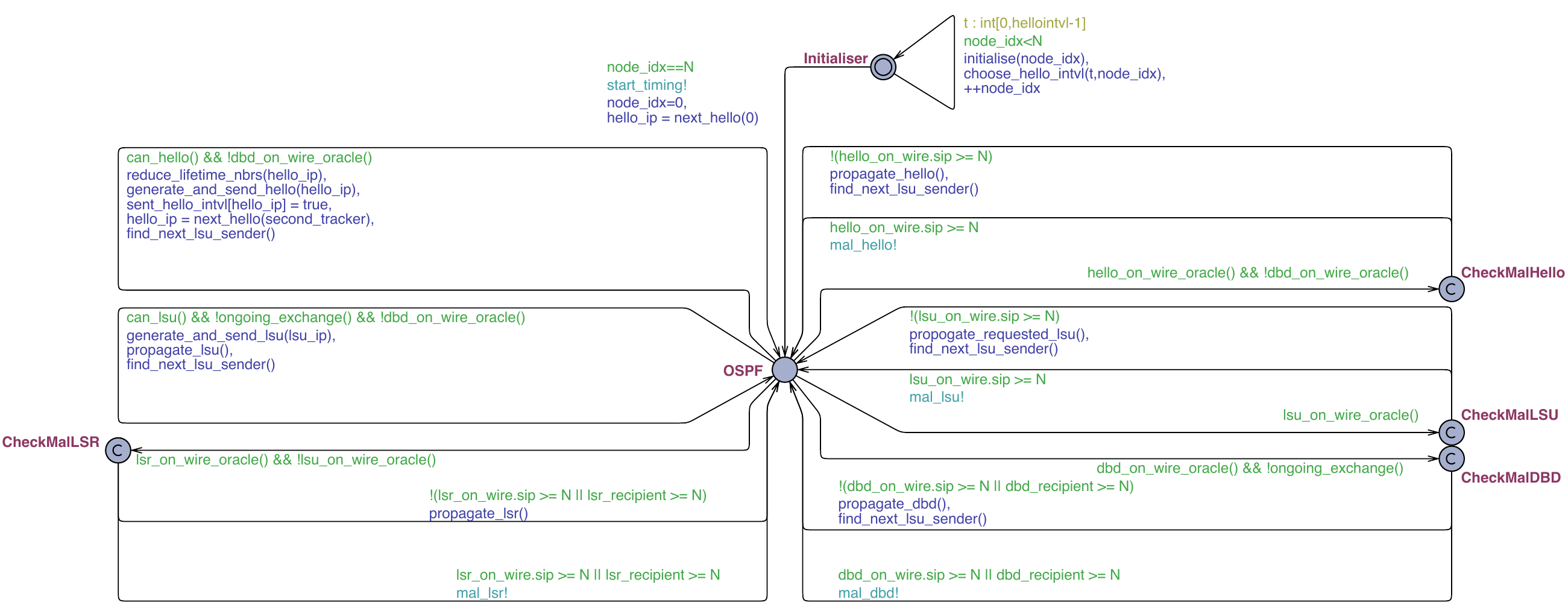}%
 \vspace{-2mm}
 \caption[]{A single automaton for OSPF, modified for attacks$^{\ref{note1}}$\label{fig:attacker}}
 \vspace{-2mm}
 \end{figure}
Such an adversary is trivial to run in parallel with the original model of OSPF, featuring individual automata for each router (\autoref{fig:ospf}), since concrete messages exist as first-class citizens. An `all mighty' attacker in our optimised model requires some minor modifications as message contents are abstracted away. \autoref{fig:attacker} demonstrates the modified automaton; each transition for delivering a message is split into two branches. One is used for the exchange of standard OSPF messages, the other allows for the exchange of messages involving a malicious party.

It is noteworthy that the new, modified automaton decouples the OSPF protocol logic from the attacker logic and allows the same automaton of OSPF to be used for any number of different attack models without modification. 
It also allows the possibility of executing standard OSPF without any malicious nodes -- in this case, our new automaton behaves identical to the original model.

\subsection{Attack Goals}
We model an attack goal as a state in the system. 
The reachability of this state determines if the attack is successful. 
Of course, it is not required that the attack is successful under all possible scenarios.

The formula \textsf{lsdp[sip][oip]} 
describes \textsf{sip}'s current view of the network topology around node \textsf{oip}, \ie 
it lists, among other information, all neighbours of \textsf{oip} that are known to \textsf{sip}. 

\paragraph{Blackhole Attack}
In this type of attack, the attacker tries to convince one or more nodes to send (data) packets via the attacker so that the packets can be dropped. 
Any packet routed through the malicious node will suffer from partial or total data loss.
These routes can only be established if either nonexistent links are installed in a node's LSDB or if some links are removed.
Both existence and removal is characterised by a malicious/fake topology:
\vspace{-1.5ex}
{\small\begin{align}\label{property1}
\begin{array}{r@{}l}
 \textsf{E\F(}&\displaystyle{\bigwedge}_{\genfrac{}{}{0pt}{}{\textsf{sip}\in S}{\textsf{dip}\in T}}\textsf{lsdb[sip][dip].malicious)} 
\end{array}
\end{align}}%
for some sets $S$ and $T$.
In case one does not want to encode non-optimal routes via concrete LSDB entries, 
one can calculate the shortest routes from a given LSDB.

\paragraph{Sub-optimal Routes}
Establishing sub-optimal routes means that the number of hops of the established route is greater than the actual shortest path. 
This attack does not cause packet loss but wastes network resources.
As packet routing is based on shortest paths, we can use the same formulas (Eq.~\eqref{property1}) to encode this attack.

\paragraph{Adjacency Interruption} The first two attacks are concerned with OSPF. An attacker can also inject or spoof OSPF messages that prevent nodes which should form an adjacency from completing the synchronisation process

\section{Initial Experiments}
Using our optimised model, we analyse the above attack goals on very small topologies. 
These experiments are not intended to be a systematic analysis of vulnerabilities, but 
are merely intended to show the power of our approach. 

\subsection{Blackhole Attack and Sub-optimal Routes}
We report on a number of examples of networks up to size 4, a full, systematic analysis on all topologies up to a much larger size is part of future work.  Some topologies used for the experiments are depicted in \autoref{fig:topologies}. 
The solid lines indicate the real topology, the dashed lines stand for the malicious topology.
In all cases, the malicious router is node $3$, and its (main) victim is node $0$. 

Attacks whose effects get subsequently corrected  by the mechanisms of the protocol, in our case OSPF, 
are called \emph{nonpersistent}. 
They can, nevertheless, temporarily affect the way the traffic is routed and can slow down the network~traffic. 

The attacks we discover rely on placing malicious entries in the victims' LSDBs. 
Each experiment related to an attack goal is run in two instances (cf.\ \autoref{ssec:cap}): 
(a) the attack starts before the network is up and running, \ie all nodes in the network 
start with an empty LSDB and the OSPF runs through its initialisation, and 
(b) the attack starts after OSPF has been initialised, \ie all LSDBs of all honest nodes contain all truthful information about the topology. 
In both cases, the attacker starts with an LSDB having malicious version of the topology as content.
This LSDB will not change during an attack.
The adversary then tries to propagate this false information through the network. 
\setlength{\tabcolsep}{.01\linewidth}
\begin{figure}[t]
  \vspace{-3mm}
\centering
\hfill
\includegraphics[width=.15\linewidth]{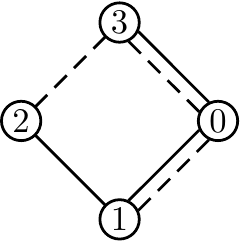}
\hfill
\includegraphics[width=.15\linewidth]{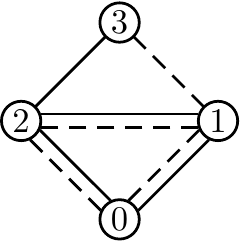} 
\hfill{}
  \caption[Network topologies:]{(a) Blackhole\ \ \ \ \ \ \ \ \ \ \ \ \ \ \ \ \ \ \  \ \ \ \  (b) Sub-optimal routes
  \label{fig:topologies}}
  \vspace{-3mm}
\end{figure}

In the blackhole attack, the adversary convinces its victim to redirect the traffic through the attacker node.
In the simple topology of \autoref{fig:topologies}(a) an adversary can be successful, \ie the property 
$\textsf{E\F (lsdb[0][2].malicious \&\& lsdb[0][3].malicious)}$ holds.
This is because most of the nodes are one-hop neighbours and proper calculation 
of shortest paths are not needed. 
 In this situation the attack is successful if node $0$ believes that 
 there is a connection between nodes $2$ and $3$ (and no connection 
 between $1$ and $2$). Instead of sending the traffic through node $1$, 
 node $0$ will then forward the traffic for node $2$ through node $3$, which will drop it. 
 
Sub-optimal routes can be installed in many topologies, such as the one depicted in \autoref{fig:topologies}(b). As for the previous attack we can simplify the attack 
goal and use the formula 
$\textsf{E\F (lsdb[0][1].malicious \&\&}$ $\textsf{lsdb[0][3].malicious)}$.
The attacker convinces node 0 that there is a connection between nodes $1$ and~$3$. 
Node $0$ then sends traffic for node $3$ through node $1$ (instead of sending 
it through node $2$). 
Node $1$ correctly believes that there is no link between itself and node~$3$, so it passes traffic through node $2$. 

The network easily recovers from these attacks by receiving further \hellomsgs, 
or by receiving further \lsumsgs sent to correct faulty information. 
Remember that these simple attacks are designed to test our setup, not to do  a full-scale analysis of vulnerabilities.

\subsection{Tear Down Adjacencies\label{sec:attacksII}}
Similarly to our use of the abstract model of OSPF, we perform initial experiments with the detailed model for adjacency building.
As before we use a nondeterministic automaton, which can inject arbitrary messages. 
We also use more limiting attack capabilities that can only inject a certain type of OSPF 
message. The one for \hellomsgs is depicted in \autoref{fig:adjacencyattacker}.

\renewcommand{\ni}{\textsf{n1}\xspace}
\newcommand{\nii}{\textsf{n2}\xspace}
\begin{wrapfigure}[9]{r}{.3\textwidth}
	\centering
	\vspace{-5mm}
	\mbox{}\hspace*{-0.5mm}\includegraphics[width=.3\textwidth]{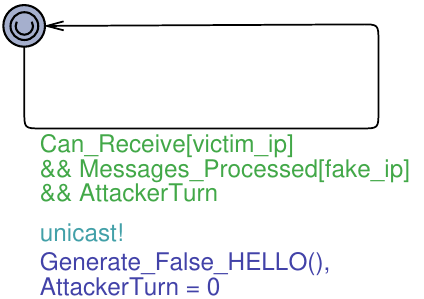}
	\vspace{-8mm}
	\caption{Injecting {\scshape Hello}s \label{fig:adjacencyattacker}}
\end{wrapfigure}
Our analysis reveals that adjacencies can be torn down. We sketch this attack; here \ni and \nii are the two routers involved.
\begin{itemize}[nosep]
	\item Sending a \hellomsg on behalf of \ni (\nii) which does not carry the IP address of \nii (\ni)  in its payload resets the adjacency building process by reverting the neighbour state back to \Init.
	\item Sending an \lsrmsg on behalf of \ni or \nii which requests an \lsadv that does not exist in the recipient's LSDB reduces the neighbour state to \ExStart if it is in a higher state at the time of  receiving. In case the neighbour state is \Down, \Init or \ExStart this injection does not have any effect.		
\end{itemize}

\begin{itemize}[nosep]
	\item Sending a \dbdmsg on behalf of \ni or \nii which meets one of the following criteria will regress to the neighbour state to \ExStart if it is in a higher state at the time of  receiving:
		\begin{itemize}
			\item The \textsc{Dbd} sequence number, a unique number which is part of every \dbdmsg, does not match the expected sequence number.
			\item The \textsc{Dbd}'s master flag is inconsistent with the current connection.
			\item The message has the init flag set -- this flag indicates that this \textsc{Dbd} is an attempt to \emph{initiate} a neighbour relationship.
		\end{itemize} 
\end{itemize}
Once a node's neighbour state has been reverted to \Init or \ExStart, it is no longer able to share information with its neighbour.
As a consequence, the adjacency needs to be re-established.
One of these attacks can delay the establishment of adjacencies;
repeating one of them indefinitely avoids the establishment of adjacencies entirely.
When checking \textsf{A\F(n1.NeighbourState == FULL \&\& n2.NeighbourState == FULL)} \uppaal generates a trace that falsifies the property,
characterising a specific attack. 

The above trace demonstrates that adjacency building can be interrupted, but it does not show  that  it is possible to tear down full adjacencies.
In order to demonstrate this, we make very minor adaptations to our model which enable the nodes to start in state \Full with their LSDBs synchronised. 
Using the same attacker automata, we use \uppaal to show that it is possible to regress our node's neighbour state to \Init or \ExStart. 
Since \uppaal does not allow for nested CTL expressions, it is not possible to verify that the nodes are unable to re-establish a state \Full. 
However, once the nodes have regressed to  \Init or \ExStart, we are able to apply our earlier  findings, as this is the same configuration as in our first experiment.

\section{Related Work\label{sec:related}}

Some of the attacks presented in \autoref{sec:attacksII} are already mentioned in~\cite{JonesLeMoigne06}. 
That analysis is based on a manual inspection of the specification of OSPF. 
While it is an impressive piece of work, our analysis is more systematic and reveals more vulnerabilities w.r.t.\ adjacency building.

As for OSPF in general, we are aware of only three other formal approaches that analyse the OSPF routing protocol, using formal methods.
We believe that our analysis is built on the most detailed, open-source formal model of OSPF~\cite{DHW20}.

Nakibly et al.~\cite{NakiblyEtAl14} have created their own model of OSPF for the model checker CBMC~\cite{CBMC}, a bounded model checker 
accepting a simplified C-language as input. 
Their model uses a fixed topology --- similar to our model-checking models --- and abstracts away from many details, including \textsc{Hello} and \dbdmsgs. The model that is available online analyses a topology with three nodes, but not more. 
Moreover, adjacency building is not modelled.
With their level of abstraction, it is not possible to model and find some of the attacks described in this paper. 

Another model, which can be fed into the Z3-Solver~\cite{Z3}, is described in~\cite{Malik}. 
The authors claim that it is a detailed model covering concepts such as designated routers --a concept we abstract from-- 
and that topologies up to $30$ nodes can be analysed. As the model is not presented in full nor available 
online, we could neither verify these claims nor compare our analysis to theirs.

\section{Conclusion and Future Work}
We have presented two advanced models of OSPF:
the first is an optimised model of an existing \uppaal-model; 
the second models adjacency building, a subprocedure 
of OSPF, in great detail. Furthermore, we have developed several attack automata that can run 
in parallel to the models of OSPF. This means our method is compositional: 
both the model and the attacker can be exchanged to variant versions. 

While this paper concentrates on our modelling efforts, future work will aim at a vulnerability analysis of OSPF at scale. 
We hope to not only rediscover well-known attacks such as the remote false adjacency attack (a.k.a.\ phantom router)~\cite{NKGB12}, which allows a malicious router to successfully establish a \emph{permanent} link to a victim, but also to discover new attacks.
We envision that we can transfer the knowledge acquired in \uppaal to (automatically) generate attacks for a real implementation.
In particular we believe that  traces produced by \uppaal can be used to generate attacks in test-bed implementations of OSPF (\eg the Quagga implementation running on the network emulator CORE~\cite{Ahrenholz08}) or in real networks.

\vspace{0.05pt}
{\myparagraph{\bf Acknowledgements:} 
This work was conducted in partnership with the Defence Science \& Technology Group 
and Data61, CSIRO, 
through the \emph{Next Generation Technologies Fund}.
We thank Jack Drury for fruitful discussions.
}
\pagebreak

\bibliographystyle{eptcs}
\bibliography{ospf}
\end{document}